\documentclass[authoryear,preprint,review,12pt]{elsarticle}
\usepackage{graphicx}
\usepackage{amssymb}
\usepackage{amsmath}
\usepackage{amssymb}
\usepackage{multicol}
\usepackage{lineno}
\usepackage[titletoc,toc]{appendix}
\usepackage[font=small,labelfont=bf]{caption}
\usepackage{longtable}

\usepackage[nolists]{endfloat}
\usepackage[inline]{trackchanges}
\addeditor{RW}

\journal{Computer \& Geosciences}
\begin{document}
\begin{frontmatter}
\title{GeoClaw-STRICHE: A coupled model for Sediment TRansport In Coastal Hazard Events}

\author[geovt]{Hui Tang\corref{cor1}}
\ead{tanghui@vt.edu}
\author[geovt]{Robert Weiss}
\cortext[cor1]{Corresponding author}
\address[label1]{Department of Geosciences, Virginia Tech, VA 24061, U.S.A.}

\begin{abstract}
STRICHE is a new model for simulating Sediment TRansport in Coastal Hazard Events, which is
coupled with GeoClaw (GeoClaw-STRICHE) to provide the hydrodynamic forcing.
Additionally to the standard components of sediment
transport models, our models also includes sediment layers and bed avalanching to
reconstruct grain-size trends as well as the generation of bed forms.
Furthermore, unlike other models based on empirical equations or sediment concentration gradient, the standard Van Leer method is applied to calculate sediment flux. We tested and verified
GeoClaw-STRICHE with flume experiment by \citet{johnson2016experimental} and
data from the 2004 Indian Ocean tsunami in Kuala Meurisi as published in
\citet{JGRF:JGRF786}. The comparison with experimental and
field data shows GeoClaw-STRICHE's capability to simulate sediment thickness
and grain-size distribution in experimental conditions. Model
results match well with the experimental and field data, especially for
sediment thickness, which builds confidence that sediment transport is correctly
predicted by this model. 

\end{abstract}
\end{frontmatter}
%\linenumbers
\section{Introduction}

Recent tsunamis, such as the 2004 Indian Ocean tsunami and the 2011 Tohoku-oki
tsunami, have caused approximately 420,000 deaths or missing reports and large
destruction of properties in coastal communities \citep{apotsos2011process,
szczucinski2012sediment}. Unfortunately, tsunamis usually have long return
period, which makes developing accurate tsunami hazard assessments very
difficult. Therefore, to assess future events, the geologic record needs to be
interrogated \citep{bourgeois1997paleotsunami, minoura2001869, jaffe2002using,
morton2007physical, Monecke2008, szczucinski2012sediment, sugawara20132011}.

Tsunami deposits are the only recorders of past events \citep{hunetal07} and
therefore play an important role in tsunami hazard assessments.  Due to the
importance of tsunami deposit, laboratory, geological and numerical modeling
studies of tsunami sediment transport have been an important \citep[and
reference therein]{tonkin2003tsunami, young2009liquefaction} but often
overlooked research area.  Most geological methods have focused on recognizing
and reconstructing the tsunami recurrence interval and characteristics of the
tsunami from geological records \citep{jaffe2002using,  burroughs2005power}.
Post-tsunami field survey can be used to interpret tsunami deposit, infer the
characteristics of tsunami flow as well as verify and validate numerical models
\citep{gelfenbaum2003erosion, borrero2005field, fritz20062004,
jaffe2006northwest, apotsos2011process, tappin2012coastal, goto2014spatial}.

As mentioned earlier,numerical modeling of tsunami sediment transport is an
important research area to improve our physical understanding of tsunami
hydrodynamics and sedimentology. Numerical modeling of sediment transport during
tsunamis can be approached in different ways. Inversion models (e.g. Moore’s
advection model, Smith's model, Soulsby’s model, TsuSedMod, TSUFLIND,
TSUFLIND-EnKF) were developed to estimate the tsunami characteristics based on
tsunami deposits \citep{Moore2007336, Smith2007362, sousby2007, jaffe2007simple,
tang2015model, wang2015inversion, tang2016tsuflind}. On the other hand, some
forward numerical models (e.g. Delft3D, Xbeach, STM, C-HYDRO3D) have been
developed to study time-varying tsunami sediment transport processes
\citep{JGRF:JGRF786, li2012tsunami, ontowirjo2013modeling, kihara2011numerical,
gusman2012numerical}. Three-dimensional models with high resolution like Delft3D
and C-HYDRO3D can be applied to incorporate vertical flow velocities and
vertical sediment concentration profiles \citep{van2004description,
kihara2011numerical}. However, three-dimension models require significant
computational resources when they are employed to consider geophysically
meaningful domain sizes \citep[e.g.,][]{sugawara2014numerical}. Therefore,
two-dimensional models are better suited. In most of tsunami sediment transport
models (C-HYDRO3D and STM), only sediment transport with uniform particle size
can be simulated \citep{kihara2011numerical, gusman2012numerical}. In order to
study sedimentary structures, mixed particle sizes and multiple sediment layers
need to be incorporated into the sediment transport models. 

With regard to the calculation of sediment flux, three major methods can be
used: empirical formulations \citep[STM,][]{gusman2012numerical}, analytical
approaches based on sediment concentration gradient \citep[Xbeach and
C-HYDRO3D,][]{roelvink2009modelling, kihara2011numerical}, and numerical models
qualified unbalance between depth-averaged concentration and equilibrium
concentration \citep[Xbeach and Deft3D,][]{roelvink2009modelling,
van2004description}. All these approaches are restricted to relatively low
sediment concentration gradient and small sediment flux condition, but they are
inappropriate to apply in situations at which high sediment concentration
gradient and large sediment flux occur. Hence, a two-dimensional, more
comprehensive and robust sediment transport model with new sediment flux
calculation method is needed. In this contribution, we present such a
two-dimensional fully coupled sediment transport model: GeoClaw-STRICHE. 

\section{Theoretical Background}
Our sediment transport model, STRICHE (Sediment TRansport In
Coastal Hazard Events), solves the governing advection-diffusion equation with a
finite volume method, add avalanching to erosion and deposition to update the
bed position. STRICHE is coupled with GeoClaw to calculate the hydrodynamics.
The standard Van Leer method, which is a widely used method in computational
fluid dynamics and aerospace engineering, is applied to calculate sediment flux.
The bed updating and avalanching scheme from \citet{roelvink2009modelling} is
used for updating topography during tsunami. STRICHE also includes
multiple grain-size classes and sediment layers to simulate sediment structure
in tsunami deposits.

\subsection{Sediment Transport Model: STRICHE}
\subsubsection{Sediment transport condition}
\label{sedcond}
The critical shear velocity is employed to decide whether sediment particles in
a given flow condition are entrained into the flow or
not. The critical shear velocity, $u_{*,cr}^b$, is estimated
with the help of an iterative procedure that discretices the Shields diagram 
\citep[see][]{Weiss2008251}. The Rouse number, $P$, is used to determine the
sediment transport condition for each grain-size class: (1) $P>2.5$: all grains
travel as bed load; (2) $1.2<P<2.5$: parts of sediment can travel as suspended
load and rest will travel as bed load (the
percentages for bed and suspended load linearly
depends on critical velocity); (3) $P<1.2$: all grains
travel in suspended condition. The critical velocity for bed load is given by:
\begin{equation} \label{eq:6} \
  U_{cr}^b=\int_{z_0}^{z}{\frac{{u_{*,cr}^b}^{2}}{K(z)}dz} 
\end{equation} where
$z_0$ is the bottom roughness from \citet{MacWilliams2005}. The eddy viscosity
profile from \citet{gelfenbaum1986} is given by: 
  \begin{equation}
    \label{eq:eddy} K(z) = \kappa \, u_* \, z \exp\left[ {\frac{-z}{h} - 3.2
    \left( \frac{z}{h} \right)^2 + \frac{2}{3} \times 3.2 \left( \frac{z}{h}
    \right)^3} \right], 
\end{equation} 
in which $z$ is the elevation above bed. The critical velocity for suspended
load is: 
\begin{equation} u_{*,cr}^s =
      \frac{w_s}{2.5 \kappa} 
\end{equation} 
and 
\begin{equation} \
    U_{cr}^s=\int_{z_0}^{z}{\frac{{u_{*,cr}^s}^{2}}{K(z)}dz}
%U_{cr}^s = \frac{w_sh^{1/6}}{2.5kgn}
\end{equation}

The combined bed and suspended load sediment
concentration is updated by solving the advection-diffusion equation
\citep{Galappatti1985}:
\begin{equation}
\label{eq4}
\frac{\partial h C}{\partial t} + \frac{\partial h  C u}{\partial x}
+\frac{\partial h  C v}{\partial y} + \frac{\partial}{\partial x}\left[D_h h
\frac{\partial C}{\partial x}\right] +\frac{\partial}{\partial y}\left[D_h h
\frac{\partial C}{\partial y}\right] = \frac{h C_{eq} - h C}{T_s}
\end{equation}
in which $C$ is the depth-averaged sediment concentration, $D_h$
denotes to the sediment diffusion coefficient, $T_s$
refers to the adaptation time, and $C_{eq}$ is the equilibrium
sediment concentration. 

\subsubsection{Transport formulations}
Two methods to calculate the equilibrium sediment concentration methods are made
available in the current version of STRICHE. For both methods, the total
equilibrium sediment concentration should be between maximum allowed sediment
concentration, $C_{max}$ and 0. The formulae to determine the equilibrium
sediment concentration are given in the following sections. It should be noted
that both of these two methods are modified to distinguish between bed and
suspend load. The first one is modified based on Soulsby-Van Rijn equations
\citep{van1984sediment, soulsby1997dynamics}. In this method, the equilibrium
sediment concentration is calculated by:
\begin{equation}
\label{eq10}
C_{eq,s}=\frac{A_{ss}}{h}\left(\sqrt{{v_{mg}}^{2}+0.018\frac{{U_{rms,2}}^2}{C_d}}-U_{cr}^s
\right)^{2.4}
\end{equation}
\begin{equation}
\label{eq11}
C_{eq,b}=\frac{A_{sb}}{h}\left(\sqrt{{v_{mg}}^{2}+0.018\frac{{U_{rms,2}}^2}{C_d}}-U_{cr}^b
\right)^{2.4}
\end{equation}
The suspended load and bed load coefficient are calculated by:
\begin{equation}
\label{eq12}
A_{ss} = 0.012D_{50} \frac{D_*^{-0.6}}{\left(\Delta gD_{50}\right)^{1.2}}
\end{equation}

\begin{equation}
A_{sb} = 0.005h\left(\frac{D_{50}}{h\Delta gD_{50}}\right)^{1.2}
\end{equation}
where $D_*$ is dimensionless grain size class. The critical velocity for bed
load and suspended load are calculated based on the method presented in section
\ref{sedcond}. When STRICHE deals with multiple grain-size classes, the median
value of grain-size distribution, $D_{50}$, in these equations is replaced by
diameter of each grain size class. Parameter $v_{mg}$ is the magnitude of the
Eulerian velocity, and $U_{rms}$ denotes the root-mean-squared velocity obtained
from linear wave theory. For the drag coefficient $C_d$ is given by:
\begin{equation}
\label{eq15}
C_d = \left(\frac{0.40}{ln\left( \frac{max(h,10z_0)}{z_0}\right)-1}\right)^2
\end{equation}

The second method used to calculate the equilibrium concentration is based on
Van Thiel-Van Rijn equations \citep{van2007unified,van2009dune}:
\begin{equation}
\label{eq16}
C_{eq,s}=\frac{A_{ss}}{h}\left(\sqrt{{v_{mg}}^{2}+0.64{U_{rms}}^2}-U_{cr}^s \right)^{2.4}
\end{equation}
\begin{equation}
\label{eq17}
C_{eq,b}=\frac{A_{sb}}{h}\left(\sqrt{{v_{mg}}^{2}+0.64{U_{rms}}^2}-U_{cr}^b \right)^{1.5}
\end{equation}
The suspended load and bed load coefficient are calculated with:
 \begin{equation}
\label{eq18}
A_{ss} = 0.012D_{50} \frac{D_*^{-0.6}}{\left(\Delta gD_{50}\right)^{1.2}}
\end{equation} 
\begin{equation}
A_{sb} = 0.015h\frac{\left(D_{50}/h\right)^{1.2}}{\left(\Delta gD_{50}\right)^{0.75}}
\end{equation} 

\subsubsection{Sediment settling velocity and density effect}
The calculation of the settling velocity $w_s$ is based on
\citet{hallermeier1981terminal} and \citet{ahrens2000fall}. For high sediment
concentration, the fall velocity is reduced:
\begin{equation}
w_{s,reduce} = (1-C)^{a_1}w_s
\end{equation}
in which $C$ is the total volume sediment concentration.
Exponent $a_1$ is estimated with the help of \citet{rowe1987convenient}, which depends on the
Reynolds particle number, $R$:
\begin{equation}
a_1 = 2.35 \frac{2+0.175R^{3/4}}{1+0.175R^{3/4}}
\end{equation}
\begin{equation}
R = \frac{w_sD}{v}
\end{equation}
Then bulk density of the fluid becomes a function of the water
density and the sediment that is suspended in it \citep{warner2008development}:
\begin{equation}
\rho = \rho_{water} +
\sum_{m=1}^{N}{\frac{C_m}{\rho_{s,m}}\left(\rho_{s,m}-\rho_{water}\right)}
\end{equation}
In this equation, $\rho_{water}$ is water density, and $C_m$ denotes sediment
concentration for grain-size class $m$. $\rho_{s,m}$ is sediment density for
grain-size class $m$. $N$ is the total number of grain-size classes. This
enables STRICHE to simulate sediment transport in high sediment concentration
condition and help to extend the density stratification in the future.

\subsubsection{Sediment Flux}
In STRICHE, the standard Van Leer method \citep{van1997flux} and the Monotonic
Upstream-centered Scheme for Conservation Laws (MUSCL) scheme \citep{van1979towards} are modified and
applied to calculate sediment flux. The Van Leer method is a finite volume
method that can provide highly accurate numerical solutions, especially for the
solutions include shocks, discontinuities, or large gradient. First, the left
and right states of the wave speed are determined on the cell surface
$(i+1/2,j)$ in which cell surface $(i+1/2,j)$ denotes the surface between cells
$(i,j)$ and $(i+1,j)$ by:
\begin{equation}
c^l_{i+1/2,j} = \sqrt{gh^{l}_{i+1/2,j}}
\end{equation}
\begin{equation}
c^r_{i+1/2,j} = \sqrt{gh^{r}_{i+1/2,j}}
\end{equation}
$h^l_{i+1/2,j}$ and $h^r_{i+1/2,j}$ refer to the left and right states of water
depth on the cell surface. The right and left states of the variables are
decided by the MUSCL scheme. For the first order, $h^l_{i+1/2,j} = h_{i,j}$,
$h^r_{i+1/2,j} = h_{i+1,j}$. Also the left and right states of Froude number are
given by:
\begin{equation}
{Fr}^l_{i+1/2,j} = \frac{U^{l}_{i+1/2,j}}{\sqrt{gh^{l}_{i+1/2,j}}}
\end{equation}
\begin{equation}
{Fr}^r_{i+1/2,j} = \frac{U^{r}_{i+1/2,j}}{\sqrt{gh^{r}_{i+1/2,j}}}
\end{equation}
$U^{l}_{i+1/2,j}$ and $U^{r}_{i+1/2,j}$ represent the left and right states of
the depth-averaged velocity at the cell surface.  After that, the positive and
negative Froude numbers are given by:
\begin{equation}
{Fr}^{\pm}_{i+1/2,j} = \frac{1}{4}\left({Fr}^{l/r}_{i+1/2,j} \pm 1 \right)^2
\end{equation}
And then parameter $\alpha$ and $\beta$ are calculated:
\begin{equation}
{\beta}^{l/r}_{i+1/2,j} = -max \left[0,1-int\left(\left|{Fr}^{l/r}_{i+1/2,j}\right|\right) \right]
\end{equation}
\begin{equation}
{\alpha}^{\pm}_{i+1/2,j} = \frac{1}{2} \left[1 \pm
sign\left(1,{Fr}^{l/r}_{i+1/2,j}\right) \right]
\end{equation}
$int$ is a function to evaluate the integer part from the variable, and $sign$
is a function to return the value of the first variable with the sign of the
second variable. The modified Froude number is determined:
\begin{equation}
{Fr_m}^{\pm}_{i+1/2,j} =
{\alpha}^{\pm}_{i+1/2,j}\left(1+{\beta}^{l/r}_{i+1/2,j}\right){Fr}^{l/r}_{i+1/2,j}-{\beta}^{l/r}_{i+1/2,j}{Fr}^{\pm}_{i+1/2,j}
\end{equation}
Finally, the sediment flux at this cell surface is:
\begin{equation}
{q_x}_{i+1/2,j} =
C^{l}_{i+1/2,j}c^{l}_{i+1/2,j}{Fr_m}^{+}_{i+1/2,j}+C^{r}_{i+1/2,j}c^{r}_{i+1/2,j}{Fr_m}^{-}_{i+1/2,j}
\end{equation}
in which $C^{l}_{i+1/2,j}$ and $C^{r}_{i+1/2,j}$ are the left and right states
of sediment concentration on cell surface. In order to reduce non-physical
dissipation in this model, we apply sediment flux limiter in the second order
simulation. Then sediment flux in y direction is calculated with the same
method.

\subsubsection{Sediment Flux Limiter}
To reduce non-physical dissipation, limit the influence of topography change and
get high order accuracy in this model, we apply sediment flux limiter and MUSCL
scheme based on \cite{van1979towards}:
\begin{equation}
\begin{split}
C_{i+1/2,j}^l = & C_{i,j}+\frac{\varepsilon}{4}[\left(1-k\right)\Psi
_{i-1/2,j}^+\left(C_{i,j}-C_{i-1,j} \right) \\
&  +(1+k)\Psi_{i+1/2,j}^-(C_{i+1,j}-C_{i,j})] 
\end{split}
\end{equation}
\begin{equation}
\begin{split}
C_{i+1/2,j}^r = & C_{i+1,j}-\frac{\varepsilon}{4}[\left(1+k\right)\Psi _{i+3/2,j}^-\left(C_{i+1,j}-C_{i,j}\right) \\
& +(1-k)\Psi_{i+1/2,j}^+(C_{i+2,j}-C_{i+1,j})]
\end{split}
\end{equation} 
It should be noted that the sediment flux limiter limits the sediment
flux by modifying the sediment concentration. Parameter
$\varepsilon$ is the order of accuracy. $k$ is decided by the type of scheme.
The flux limiter is given by:
\begin{equation}
\Psi_{i+1/2,j}^+ = \Psi\left(r_{i+1/2,j}^+\right)
\end{equation}
\begin{equation}
\Psi_{i+1/2,j}^- = \Psi\left(r_{i+1/2,j}^-\right)
\end{equation}
Where sediment concentration gradient $r_{i+1/2,j}^+ =
\frac{C_{i+2,j}-C_{i+1,j}}{C_{i+1,j}-C_{i,j}}$ , $r_{i+1/2,j}^- =
\frac{C_{i,j}-C_{i-1,j}}{C_{i+1,j}-C_{i,j}}$. The flux limiter function, $\Psi$,
is based on \citet{van1982comparative}.

\subsection{Morphology Update}
During deposition or erosion, the surface elevation, $Z_b$, is updated by:
\begin{equation}
\frac{\partial Z_b}{\partial t}+\frac{f_{mor}}{1-p}\left(\frac{\partial
q_x}{\partial x}+ \frac{\partial q_y}{\partial y}\right) = 0 
\end{equation} 
in which $q_x$ and $q_y$ are sediment flux in horizontal direction and $f_{mor}$
is a morphological acceleration factor from \citet{reniers2004morphodynamic}. To
account for the slumping of sediment during tsunami, avalanching scheme from
\citet{roelvink2009modelling} is incorporated.  When the slope between two
adjacent grid cells exceeds the critical slope, sediment is exchanged between
these cells to reduce the slope below the critical slope.

\subsection{Sediment Setting}
\subsubsection{Sediment Classes}
In order to be able to reproduce vertical grain-size trends, different
grain-size classes need to be implemented as well as different layers of
erodible bed sediments. The grain-size distributions are represented by discrete
grain-size classes. It should be noted that natural sediment are most likely
mixtures of cohesive and non-cohesive sediments. While sediment entrainment is
affected by cohesive sediments, the influence of cohesive material on
non-cohesive sediments is not trivial. A dynamical model to take cohesive
sediments into account is still under development and will be included in future
versions of STRICHE.

\subsubsection{Sediment Layers}
\begin{figure}[!ht]
  \centering
    \includegraphics[width=0.75\textwidth]{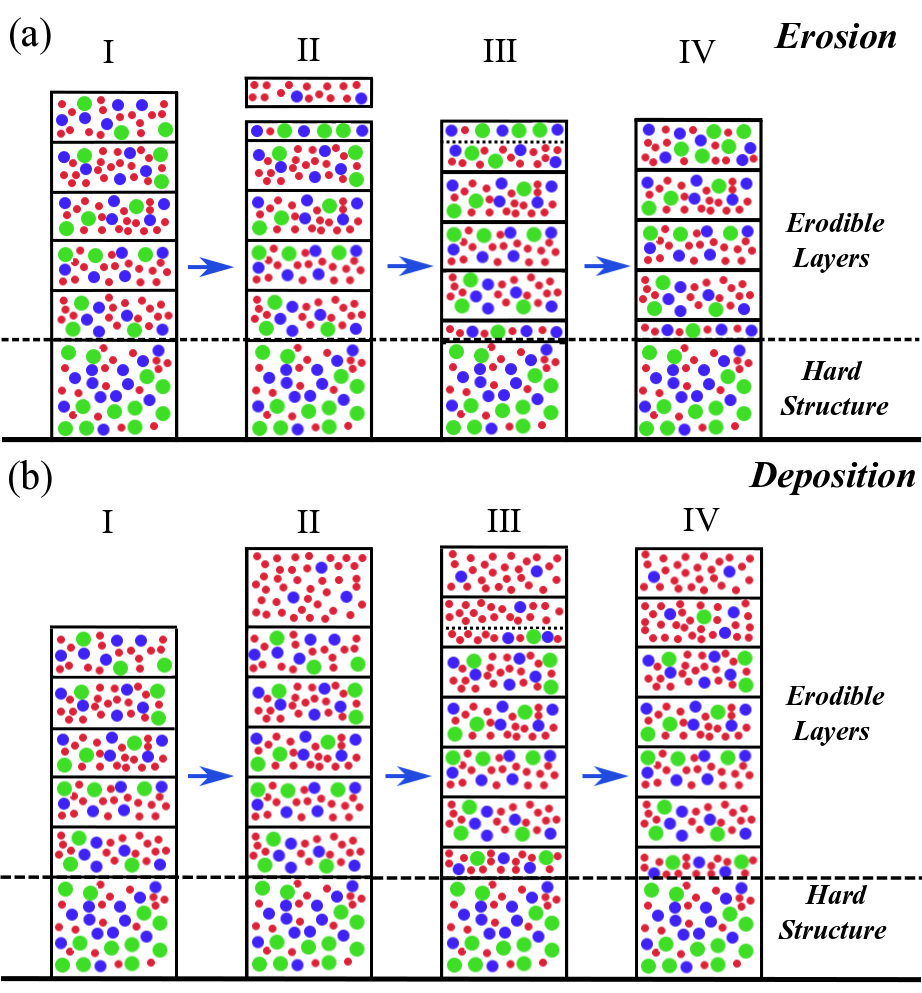}
  \caption{Concept model of sediment layers setting. The sediments are separated
  to erodible layers and hard structure. (a): Concept model for sediment layers
  during erosion; I: original sediment condition; II: flow eroded part of
  sediments; III: Remap sediment layers; IV: recalculate sediment properties for
  each layers (b): Concept model for Sediment layers during deposition; I:
  original sediment condition; II: flow deposited part of sediments; III: remap
  sediment layers; IV: recalculate sediment properties for each layers.}
    \label{fig1}
\end{figure}

The thickness of each layer of the erodible bed is user-defined, and the top one
is considered the active layer. Sediment in erodible sediment layers can be
transported by suspended load and/or bed load throughout the entire simulation.
However, only the sediment in the active layer is available for transport at any
given time step. It is assumed that in one time step not more than the thickness
of the active layer can be eroded (see Fig.\ref{fig1}a-I \& II.) After the
current time step, the layers are remapped starting with the erodible bed from
the top to assure that the top layer has full thickness available for erosion
(see Figs.\ref{fig1}a-II \& III). A similar process is implemented for
deposition as depicted in Figs.\ref{fig1}b-I \& IV.  After this remapping
procedure and after possible bed avalanching, the grain-size distribution in
each layer are recalculated. 

\subsection{Hydrodynamic Model: GeoClaw}
For simulating hydrodynamics of tsunami waves, GeoClaw is employed. GeoClaw is
based on the Clawpack software. GeoClaw solves the nonlinear shallow water
equations with high-resolution shock capturing finite volume method
(Godunov-type method) on logically rectangular grids \citep{LeVeque2011}.
GeoClaw also features adaptive mesh refinement (AMR) to achieve the efficient
computations of large-scale geophysical problems such as tsunamis and storms.
The algorithms and theories applied in GeoClaw are discussed with more details
in \citet{LeVeque2011}. GeoClaw is verified and validated against analytical
solutions and real cases presented in \citet{gonzalez2011} and
\citet{Arcos2015}. Furthermore, GeoClaw has been used in a number of tsunami
studies and other applications \citep{Gavin4808, George2298, Mandli201436,
LoyceM2015}.

\subsection{Model algorithm}
\begin{figure}[!ht]
  \centering
    \includegraphics[width=0.7\textwidth]{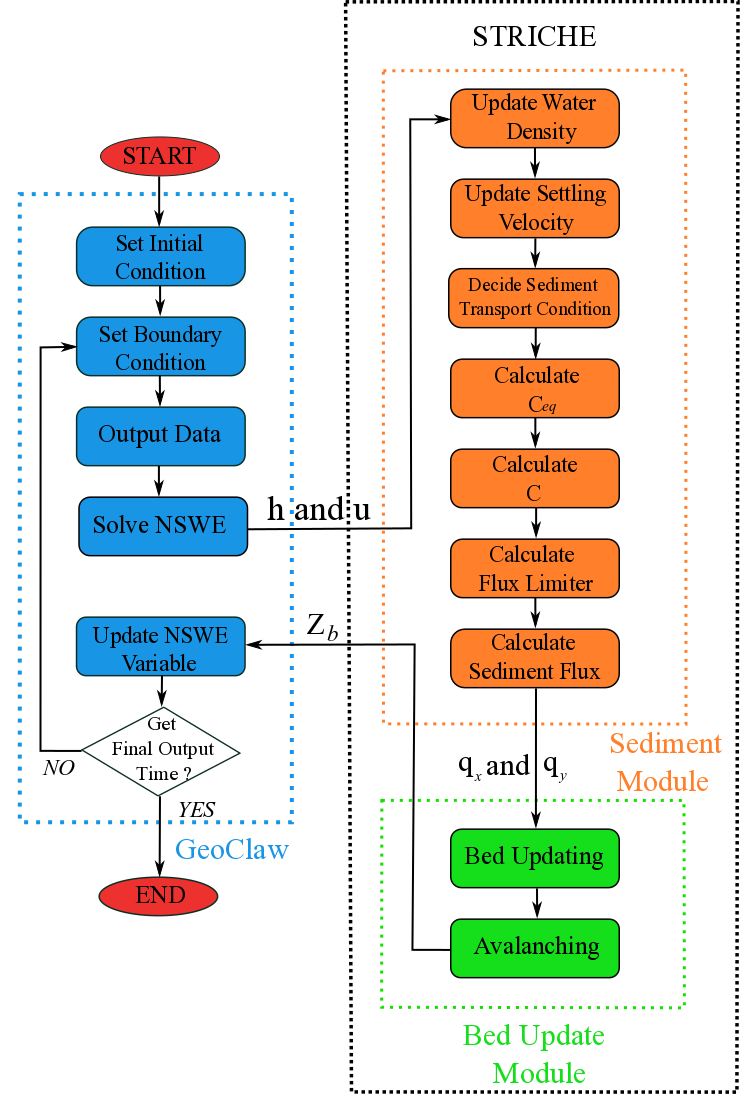}
  \caption{Flowchart for model algorithm}
  \label{fig2}
\end{figure}

For the coupling between GeoClaw and STRICHE, a two-way coupling but separately
solving system is utilized. The hydrodynamic model, sediment transport model,
and morphodynamic model are constructed as three separate modules in
GeoClaw-STRICHE. STRICHE includes the sediment transport model and morphodynamic
model (Fig. \ref{fig2}). In each time step, the hydrodynamic model (GeoClaw)
outputs hydrodynamic condition to the sediment transport model. The sediment
transport model passes sediment fluxes to the morphodynamic model. In turn,
STRICHE returns topography information to GeoClaw and change the source term of
shallow water equations (Fig. \ref{fig2}).

\section{Model Validation}
In this section, we provide two validation cases that highlight the capabilities
of coupled model GeoClaw-STRICHE. The first case is a flume experiment of
tsunami sediment transport processes in an open channel. This case demonstrates
the ability of GeoClaw-STRICHE to reproduce sediment thickness and grain-size
distribution in experimental conditions. The second case is a real application
with complex topography from the 2004 Indian Ocean tsunami. This case
demonstrates the capability of GeoClaw-STRICHE to simulate sediment thickness in
a realistic application with complex bathymetry.

\subsection{Flume Experiment Case}
\begin{figure}[!ht]
  \centering
    \includegraphics[width=0.8\textwidth]{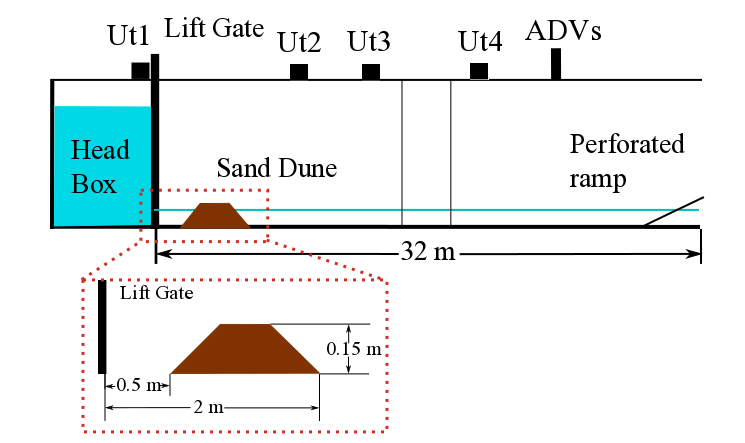}
  \caption{Schematic diagram for experiment setting with major components shown
  in \citet{johnson2016experimental}. Ut: ultrasonic transducers for water depth
  measurement; ADVs: two side-looking Nortek Vectrino ADVS for flow velocity
  measurement. Sediment source was located 0.5 to 2 m in front of the lift gate
  as a sand dune about 1.5 m long and 0.15 m high. There is a
  computer-controlled lift gate at left side, perforated ramp at right side, and
  a smooth bed without slope between them.}
  \label{fig3}
\end{figure}

GeoClaw-STRICHE is tested by simulating several open channel cases from
\citet{johnson2016experimental}. These laboratory experiments originally were
conducted to evaluate the tsunami inversion model assumptions and qualify
uncertainties. These experiment cases were set in a 32 m long, 0.5 m wide and
0.8 m high water tank at the University of Texas at Austin (Fig. \ref{fig3}).
There was a smooth bed without slope and a lift gate to control input flow. The
first three cases ran with the same grain-size distribution of the sediments
(Fig. 4a, source 1) and three different initial water depths in the water tank
(0, 10 and 19 cm).  Case IV, V and VI set with the same initial water depth (8
cm) and three different grain-size distributions in sediment (Fig. 4a). The
sediment was located between $0.5 m$ and $2.0 m$ from the lift gate, and had a
trapezoidal cross section that was $1.5 m $ long and $0.15 m$ high (Fig.
\ref{fig3}). When the gate was lifted, the flow eroded the sand dune like a
tsunami would erode a coastal dune during propagation.

Water depth was measured at three locations (Ut2: 7.25 m, Ut3: 17.7 m and Ut4:
19.1 m from the lift gate in Fig. \ref{fig3}) by using ultrasonic transducers.
Flow velocity was measured by two side-looking Notrek Vectrino ADVs located 19.3
m from the lift gate (Fig. \ref{fig3}). We calculated mean value and smoothed
the original data including water depth and velocity to make them easy to
compare with simulation data. After sediment deposition and water drainage,
sediment thickness was measured every 25 cm. Grain-size distribution was
measured every meter at 1/8 $\Phi$ resolution by an imaged-based Retsch
Camsizer. The sediment samples used to analyze grain-size distribution included
the entire thickness of deposit in the sample location.

\begin{figure}[!ht]
  \centering
    \includegraphics[width=0.95\textwidth]{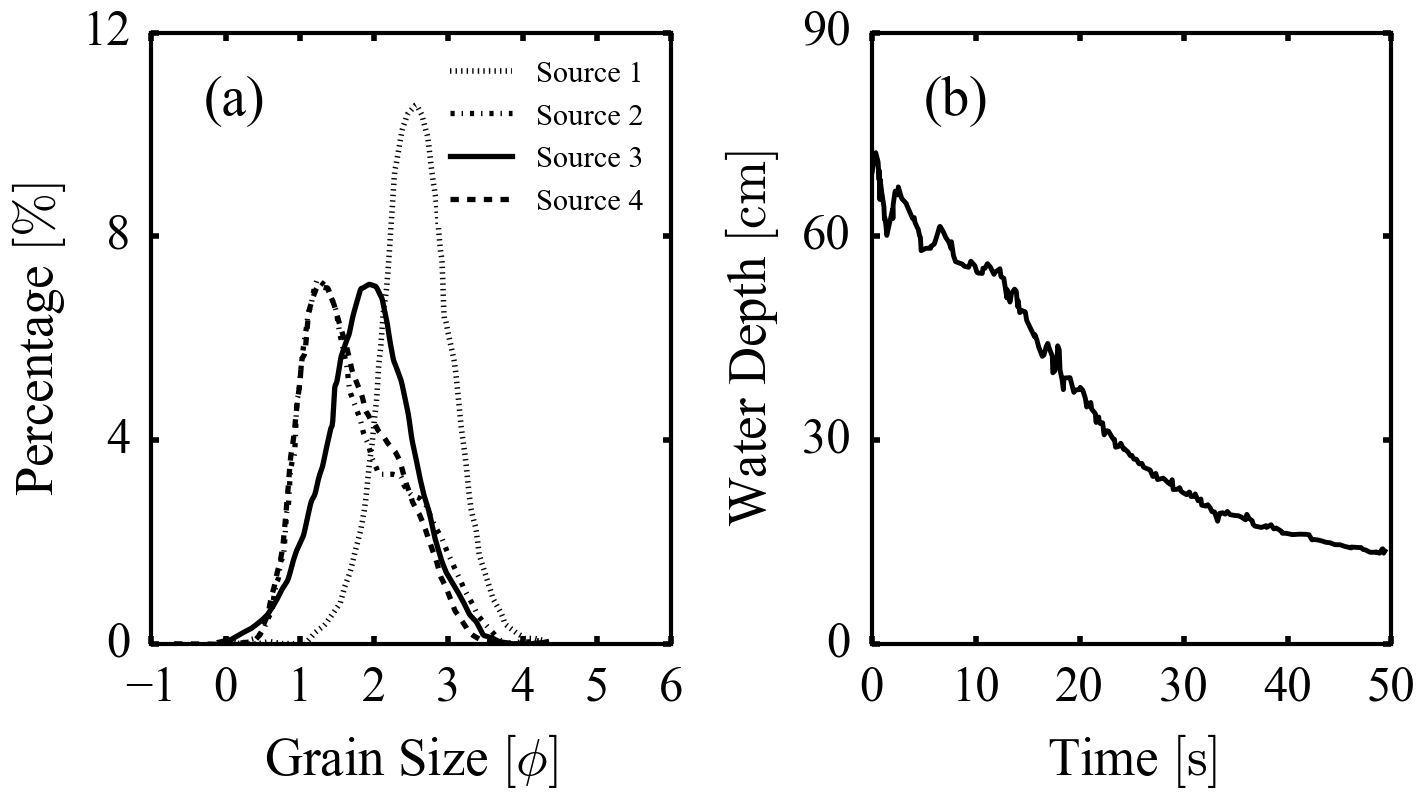}
  \caption{Initial setting for experiment and model based on \citet{johnson2016experimental}: (a): Grain-size distributions of sediment source (source 1-4); (b): Water depth measure at headbox and boundary condition in simulations.}
  \label{fig4}
\end{figure}

For this application, 2D depth-averaged simulations including bed load and
suspended load by GeoClaw-STRICHE were carried out. The water depth at Headbox
(Fig. \ref{fig4}b) was treated as a boundary condition in simulations. All
simulations started with a sand dune located within $0.5 m$ to $2.0 m$ m from
the gate (Fig. \ref{fig3}). Grain-size distributions of sediment were
discretized with ten grain-size classes. The grain-size distributions for the
experiment are depicted in Fig. \ref{fig4}a. The sediment layer in this case
will be only separated to one erodible layer and hard structure in the model
runs. The computational domain is $25 \times 1600$ with mesh size $0.02 m\times
0.02 m$.

\begin{figure}[!ht]
  \centering
    \includegraphics[width=0.9\textwidth]{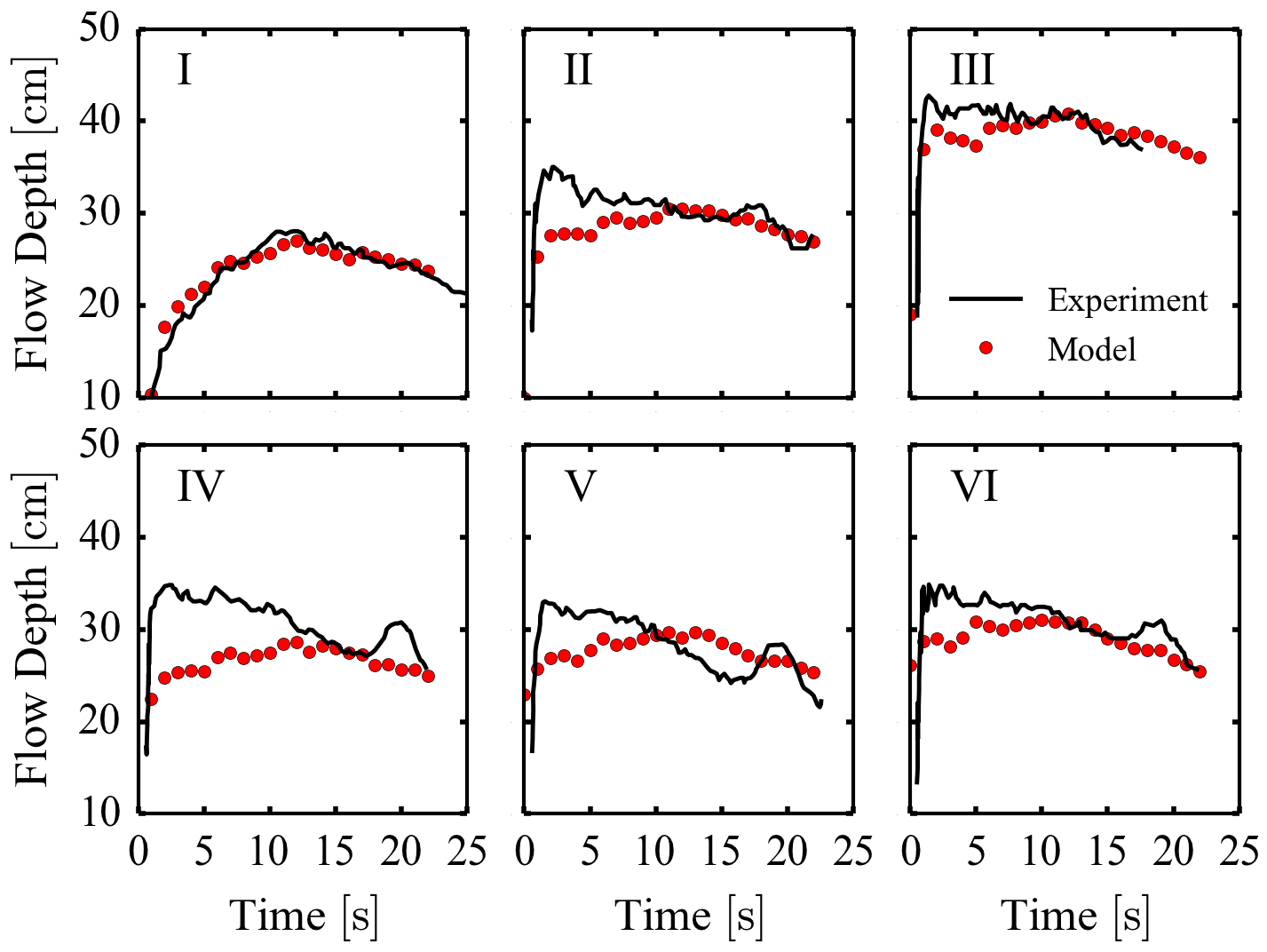}
  \caption{Measured flow depth (black line) and model results (red circle). I:
  source 1 on dry land; II: source 1 in 10 cm water; III: source 1 in 19 cm
  water; IV: source 2 in 8 cm water; V: source 3 in 8 cm water; VI: source 4 in
  8 cm water.}
  \label{fig5}
\end{figure}
The flow depth was measured in the experiments by an ultrasonic transducer at
17.7 m downstream from the lift gate, with the exception of case III, where the
flow depth was measured from sidewall with the help of video frames analysis.
The experimental results from \citet{johnson2016experimental} show that the
shallowest water depth achieved from the case I (Fig.  \ref{fig5}-I), and
deepest one achieved from case III (Fig. \ref{fig5}-III). The largest water
depth in all cases, except case I, almost appeared immediately after bore front
arrival (Figs. \ref{fig5}-II to \ref{fig5}-VI). Flow depth results from case IV
to VI are very similar in the experiment as shown in Figs.  \ref{fig5}-IV to
\ref{fig5}-VI. For all cases, the water depth results from GeoClaw-STRICHE are
in good agreements with the measurements, especially for case I. The water depth
simulation results at the first ten seconds are slightly underestimated compared
with experiment results for case II to VI (Figs.  \ref{fig5}-II to
\ref{fig5}-VI). The water depth results from 10 to 20s for case I, II, III and
VI are well captured in the simulation (Figs. \ref{fig5}-I, II, III, VI). At the
end of the simulation, calculated water depths slightly deviate from the
measurement data in case IV to VI (Figs. \ref{fig5}-IV to VI).

\begin{figure}[!ht]
  \centering
    \includegraphics[width=0.9\textwidth]{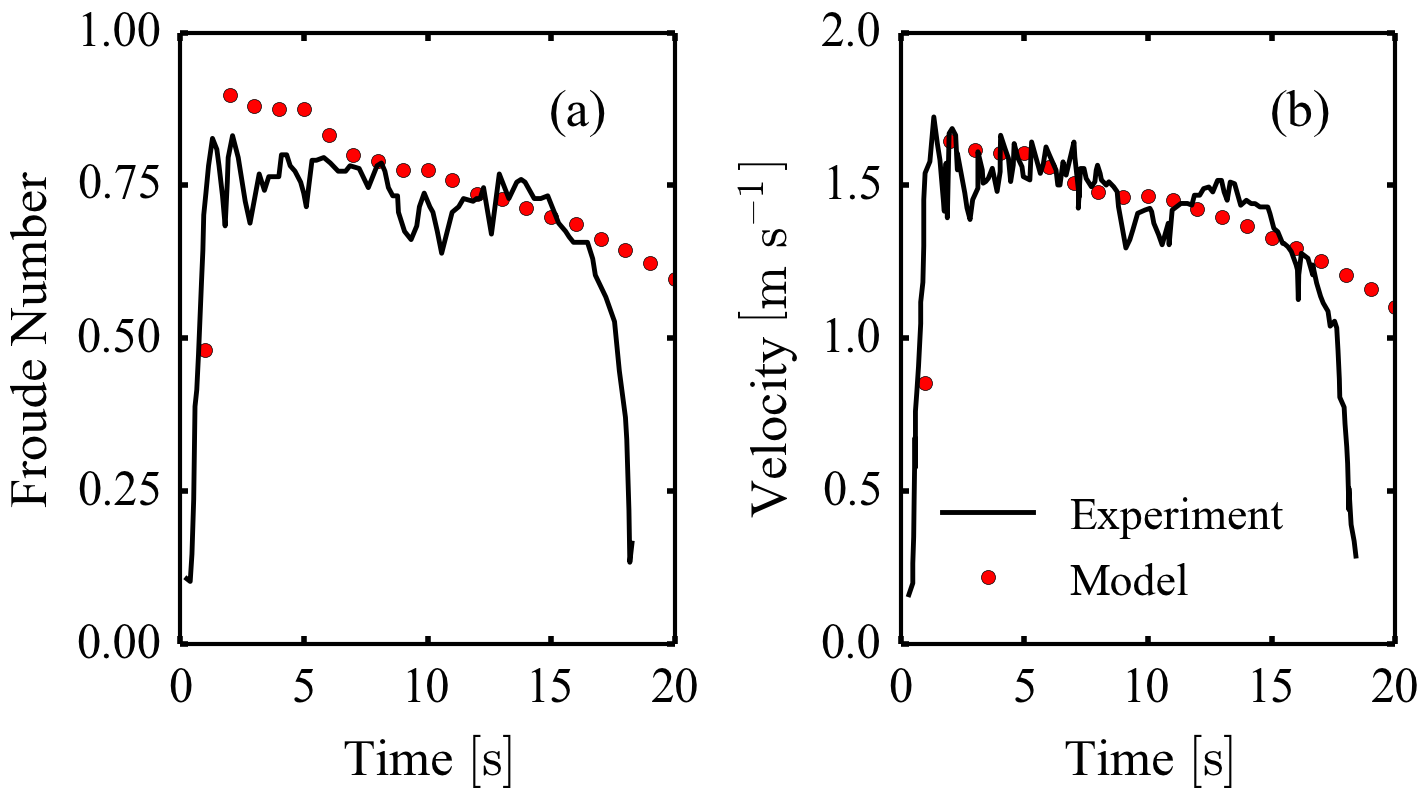}
  \caption{ (a): Froude number from experiment in case III (black line) and
  model results (red circle). (b): Flow velocity from experiment for case III
  (black line) and model results (red circle).}
  \label{fig6}
\end{figure}

Flow velocities were measured by ADVs at $19.3 m$ from the gate. As there is
little difference between velocities measured at 9 cm and 15 cm above the bed,
the velocities measured in the experiment are considered as depth-averaged
velocity \citep{johnson2016experimental}. It should be noted that the Froude
number for case III is based on the flow depth at 18.7 m downstream. The flow
velocity was highest right after the bore and then decreased gradually
(Fig.\ref{fig6}b).  The model slightly overestimates the Froude number in the
first five seconds of case III (Fig.  \ref{fig6}a). Also the model results
follow a similar pattern, but there is an obvious overprediction after 15s for
both velocity and Froude number (Fig. \ref{fig6}).

\begin{figure}[!ht]
  \centering
    \includegraphics[width=1.0\textwidth]{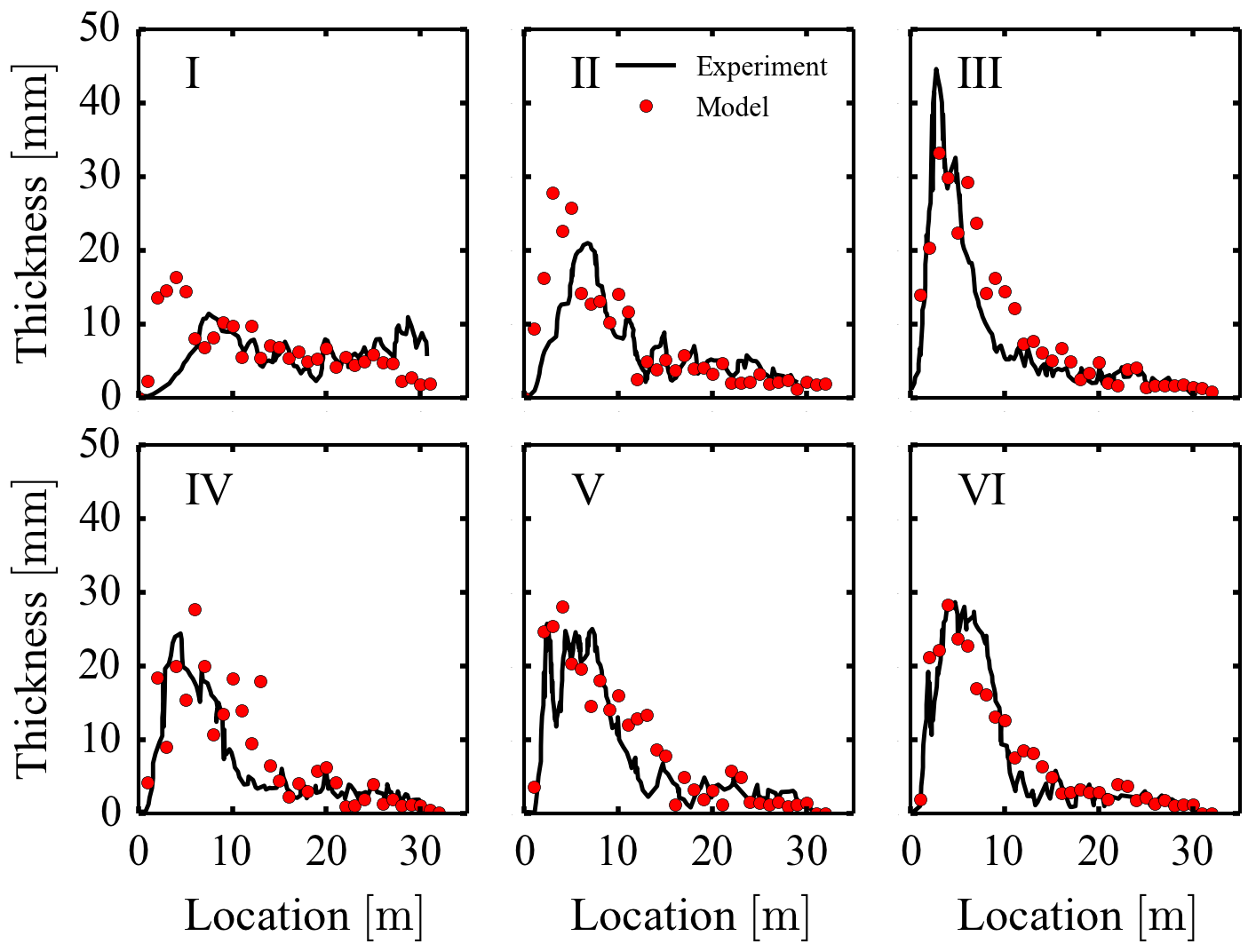}
  \caption{Sediment thickness from experiment (black line) and model results
  (read circle). I: source 1 on dry land; II: source 1 in 10 cm water; III:
  source 1 in 19 cm water; IV: source 2 in 8 cm water; V: source 3 in 8 cm
  water; VI: source 4 in 8 cm water.}
  \label{fig7}
\end{figure}

Figure \ref{fig7} compares measured and calculated sediment thickness. In case
I, most of the sediment dune was eroded by the flow (Fig. \ref{fig7}-I). For the
rest of cases, the sediment thicknesses increased and then decreased. The
thickness trends are not significantly different for case IV to case VI in both
experimental and model results (Figs. \ref{fig7}-IV to VI). In most cases, the
dune was significantly modified, and only a very small sediment wedge was left
behind (Figs. \ref{fig7}-IV to VI).  A significant downstream thinning of the
sediment layer was observed in cases II to VI (Figs. \ref{fig7}-II to VI). There
is again a fairly good agreement for sediment thickness between simulation and
experiment in all cases (Fig.  \ref{fig7}).  The thickest sediment thickness
generally occurred between $1 m$ and $6 m$ from the gate (Fig. \ref{fig7}). 

\begin{figure}[!ht]
  \centering
    \includegraphics[width=0.9\textwidth]{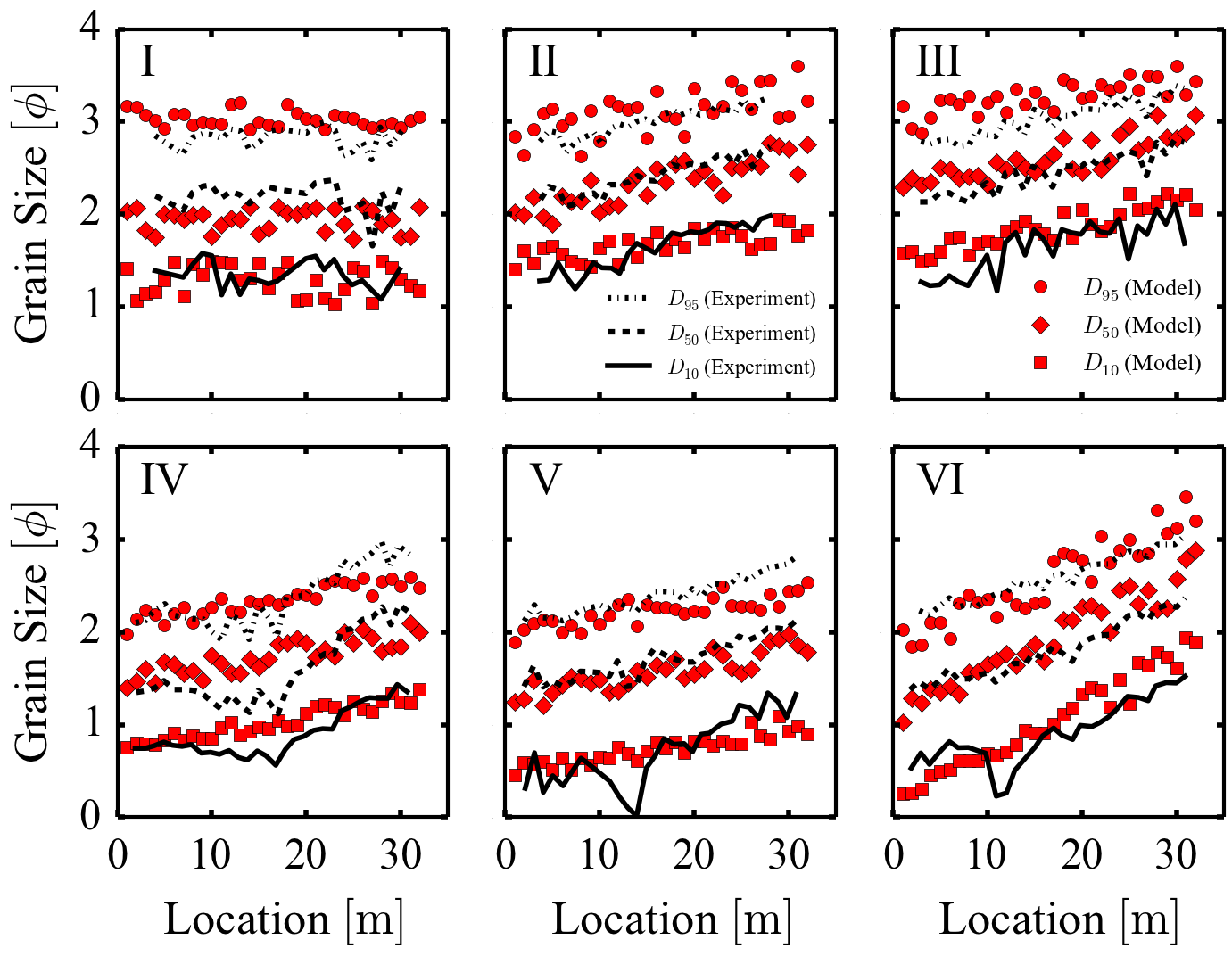}
  \caption{D10, D50, D95 from experiment (line) and model results (marker). I:
  source 1 on dry land; II: source 1 in 10 cm water; III: source 1 in 19 cm
  water; IV: source 2 in 8 cm water; V: source 3 in 8 cm water; VI: source 4 in
  8 cm water.}
  \label{fig8}
\end{figure}

Figure \ref{fig8}  shows the measured and simulated $D_{95}$, $D_{50}$ and
$D_{10}$. The trends of $D_{50}$ along the tank are very similar between
experimental and simulated results for all cases. The $D_{50}$ decreases from
initial dune position towards the end of tank (Fig. \ref{fig8}). However, the
grain-size fining trend is not very obvious in case I (Fig. \ref{fig8}a-I).
Without loss of generality, model results for $D_{50}$ fit well with
experimental results. The grain-size trends are very similar for $D_{95}$ and
$D_{10}$ in case II and case III (Figs. \ref{fig8}-II \& III). The model can
reproduce the $D_{95}$ and $D_{10}$ for these two cases.  For cases V and VI,
there is significant underprediction for $D_{95}$ and $D_{10}$ in simulations,
especially for $D_{10}$ from 10 m to 20 m (Figs. \ref{fig8}-V \& VI).

\subsection{The 2004 Indian Ocean Tsunami Case}
\begin{figure}[h!]
  \centering
    \includegraphics[width=1.0 \textwidth]{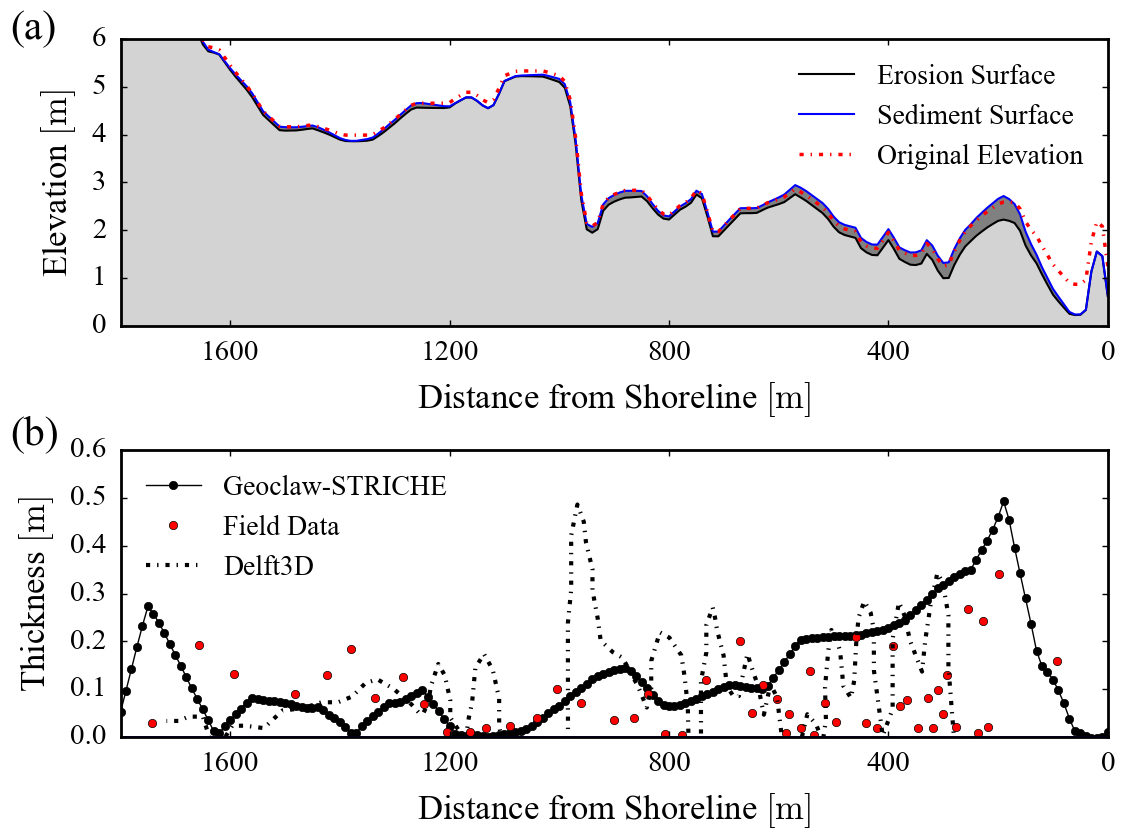}
  \caption{ (a): Maximum erosion surface, final sediment surface and original
  surface in study transect for the 2004 Indian Ocean tsunami in Kuala Meurisi;
  (b): Model results, field data and model results from Delft3D based on
  \citet{JGRF:JGRF786}.}
  \label{fig9}
\end{figure}

Validation data retrieved in a controlled environment does not exist for
complicated topographies. Therefore, field data play a crucial role in model
validation. In this section, a real case from Kuala Meurisi after the 2004
Indian Ocean tsunami is presented. Nearshore bathymetry, onshore topography and
incident wave characteristics were taken from \citet[][; red dotten line in Fig.
\ref{fig9}a represents the pre-tsunami onland topography]{JGRF:JGRF786}.  A grid
spacing of 10 m is used in this simulation. The erodible sediment layer is 5 m
thick with five grain-size classes (-1 to 4 in $\phi$ scale) with equal
percentages. 

The GeoClaw-STRICHE results are compared with the field data from Kuala Meurisi
and respective model results computed with Delft3D in Fig. 9. Figure \ref{fig9}a
shows the maximum erosion surface, final sediment surface and original sediment
surface.  Figure \ref{fig9}b compares the sediment thickness along the cross
section measured in the field and calculated with Delft3D \citep{JGRF:JGRF786}
with the results from GeoClaw-STRICHE.  The results from GeoClaw-STRICHE show
similar trends with the field observations. Both simulation results
(GeoClaw-STRICHE and Delft3D) and field data show significant erosion in first
100 meters. In some places, such as at 150 m inland, the final sediment surface
is below the original surface, but yet there is some sediment deposited, which
is a model result that fits with field observation. From 200 m to 600 m inland,
the model results are imperceptibly overestimated. The model results and field
data fit well from 600 m to 1300 m inland. After 1300 m some underestimations
appear in this area again. Compared with GeoClaw-STRICHE results, Delft3D
results seem to have more fluctuations and overestimations from 200 m to 1000 m
inland.

\section{Discussion}
\subsection{Interpretation of test case results}
With the help of GeoClaw-STRICHE, we can reproduce the fluid conditions as well
as sediments distributions in experimental cases and real-world tsunamis.
Figures \ref{fig5} to \ref{fig8} summarize our simulation results for the
experiments.  For all cases, especially for case I, model results are in very
good agreement with the experiments (Figs. \ref{fig5}, \ref{fig7}, \ref{fig8}).
The slight difference of the flow depth between model and experiment results at
the first 10s for case IV to VI can be explained by the sand dune development
under unsteady flow. There are obvious overpredictions after 50s for both
velocity and the Froude number in Case III (Fig. \ref{fig6}). While this fact
can arguably be ascribed to the imperfect model, we note that it is also
possible that the data suffer a large uncertainty due to the fact that ADVs can
return erroneous data if the flow is really shallow or a large number of air
bubbles are present. Note that the Froude number in this case is calculated
based on 18.7 m's flow depth and 19.1 m's velocity, which may also contribute to
the overprediction. The thickness trends are not significantly different between
case IV to VI in both experimental and model results, which indicates the
grain-size distribution did not significantly control the thickness distribution
for this experimental condition. However, sediment transport in the dry land
condition seems to be much more difficult to simulate compared to other cases
due to strong turbulence. As a result the calculated thicknesses are deviated
from experiment results in Fig. \ref{fig7}-I. The $D_{10}$ contains apparent
difference in case V and VI between observation and model results (Fig.
\ref{fig8}). It is likely that underestimations in the fluid dynamics from model
have more influence on large particles than small ones. 

Figure \ref{fig9} shows the model results from GeoClaw-STRICHE and Delft-3D as
well as field measurement for the 2004 Indian Ocean tsunami on Kuala Meurisi.
The model results show similar trends with the field measurements compared with
results from Delft3D, which indicates that the strength of sediment transport is
correctly predicted in GeoClaw-STRICHE for the real-world tsunami. From 200 m to
600 m inland, the model results are imperceptibly overpredicted, which may be
caused by the complex topography. The fluctuations and overestimations in
Delft3D’s results may due to the different way to calculate sediment thickness.

\subsection{Model Limitations and Future works}
The GeoClaw-STRICHE has proven its capability of simulating sediment transport
by comparing experimental, field and results from Delft3D with our model.
However, there are still some limitations for the present version of
GeoClaw-STRICHE. A significant assumption of GeoClaw-STRICHE is that tsunami
sediments are represented with several grain-size classes. As a result, the
increasing number of sediment grain-size classes will significantly increase the
computational load.  Meanwhile, GeoClaw-STRICHE can only simulate sand sediment
transport. However, tsunamis can have the power to transport almost all types of
sediment from mud to boulder. Another important assumption of GeoClaw-STRICHE is
that sediment will be picked up when shear velocity is larger than the critical
shear velocity. In this model, we use Shield diagram to estimate the critical
shear velocity for simplicity. The Shield diagram can only deal with particle
size between 9 to -4 $\Phi$. Also the critical shear velocity from Shield
diagram usually is considered as a rough estimate. At the same time, the
depth-averaged velocity for sediment movement is calculated based on water
depth, shear velocity and eddy viscosity profile. The choice of eddy viscosity
profile will extremely influence the velocity estimation.

Future model improvement will be developed in an open source, community
development approach. In order to solve computational load problems, the OpenMP
and/or MPI (Message Passing Interface) should be implemented to increase
calculation efficiency. Algorithms to represent cohesive sediment, gravel, and
boulder sediment transport, as well as density stratification are under
development. We also plan to investigate alternative approaches to calculate
settling velocity, shear velocity, critical shear velocity, and sediment flux.

\section{Conclusions}
In this paper, we propose a tsunami sediment transport model, STRICHE.  STRICHE
is fully coupled with GeoClaw for computing the hydrodynamics (combined model is
referred to as GeoClaw-STRICHE).  STRICHE's features include multiple sediment
grain-size classes and sediment layers. The sediment concentration is computed
by an advection-diffusion equation from \citet{Galappatti1985}. The standard Van
Leer method is applied for calculating sediment flux. To avoid the nonphysical
model instability, a flux limiter is used. The bed-updating module controls the
topography change and the mass balance between two neighbor cells.
GeoClaw-STRICHE tracks sediment thickness and properties for each morphological
step and updates topography information for fluid dynamic module. The quality of
the matches between experimental, field and numerical results, show that STRICHE
is capable of reliably simulate sediment transport during coastal hazard events,
such as tsunamis. The coupling with other hydrodynamic models, for example to
consider storm waves and surge, will show its capabilities for general coastal
hazard flooding. Furthermore, in its current version, but especially with the
planned improvement, STRICHE is and will more so be able in the future to be
applied on only to modern cases, but also to coastal hazard flooding event in
the past where the only physical evidence is the deposits.

\newpage
\begin{appendices}

\section*{Appendix A: Symbol List}
\begin{footnotesize}
\begin{longtable}{ l l l }
\endfirsthead
\endhead
\endfoot
\endlastfoot
  Symbol    			&Dimensions					& Description													\\
  \hline 
   $A_{ss}$			        &$1$		&Suspended load coefficient                    							\\
   $A_{sb}$			        &$1$		&Bed load coefficient                    							\\
   $a_1$			    	        &$1$			    &Settling velocity reduction coefficient                							\\
  $B$				    &m						& Bathymetry information
 								\\
  $C$					&$\text{m}^{3}\text{m}^{-3}$		&Depth-averaged sediment concentration                    							\\
  $C_m$					&$\text{m}^{3}\text{m}^{-3}$		&Depth-averaged sediment concentration for grain-size class m                							\\
  $C_{eq}$					&$\text{m}^{3}\text{m}^{-3}$		&Equilibrium depth-averaged sediment concentration                    							\\
  $C_{eq,s}$			    &$\text{m}^{3}\text{m}^{-3}$		&Equilibrium suspended sediment concentration                    							\\
  $C_{eq,b}$			    &$\text{m}^{3}\text{m}^{-3}$		&Equilibrium bed sediment concentration                    							\\
  $C_{max}$			    &$\text{m}^{3}\text{m}^{-3}$		& Maximum allowed sediment concentration                    							\\
  $C_{i+1/2,j}^{l/r}$			&$\text{m}^{3}\text{m}^{-3}$ & Left or right state of depth-averaged sediment concentration  
  								\\ 
     						&							&of cell surface $(i+1/2,j)$
    \\
    $C_{m}$			&$\text{m}^{3}\text{m}^{-3}$ & Depth-averaged sediment concentration for grain-size class $m$
  								\\                              
   $C_{i,j}$			&$\text{m}^{3}\text{m}^{-3}$ & Depth-averaged sediment concentration at the cell (i,j)
  								\\ 
   $C_{i+1,j}$			&$\text{m}^{3}\text{m}^{-3}$ & Depth-averaged sediment concentration at the cell (i+1,j)
   								\\ 
      $C_{d}$			    	&$1$					&Drag coefficient                  							\\                             
    $c_{i+1/2,j}^{l/r}$			&$\text{ms}^{-1}$ & Left or right state of wave speed of cell surface $(i+1/2,j)$
   							\\ 
 $D$			    	        &$\text{mm}$			    &Grain size                 							\\ 
  $D_{50}$			        &$mm$		&Median Grain size                  							\\       
  $D_*$			    	    &$\text{1}$			    &Dimensionless grain size                 							\\
  $D_h$						&$1$					 & Sediment diffusion coefficient
  								\\                               
  $Fr_{i+1/2,j}^{l/r}$			&$1$ & Left or right state of Froude number of cell surface $(i+1/2,j)$
  								\\ 
  ${Fr_m}_{i+1/2,j}^{\pm}$			&$1$ & Plus or minus modified Froude number of cell surface $(i+1/2,j)$
  								\\ 
  $f_{mor}$			   &$\text{1}$ & Morphological acceleration factor               
                                 \\                                 
  $g$				    &$\text{ms}^{-2}$ 		& Gravity acceleration
  								\\
  $h$ 					&m						&Water depth 														\\
  $h_{i+1/2,j}^{l/r}$			&$\text{m}$ & Left or right state of water depth of cell surface $(i+1/2,j)$
  								\\ 
   $h_{i,j}$					&$\text{m}$ & Water depth at the cell (i,j)
  								\\ 
   $h_{i+1,j}$				&$\text{m}$ & Water depth at the cell (i+1,j)
  								\\ 
  $K(z)$					&$\text{m}^{2}{s}^{-1}$	 & Eddy viscosity profile
  								\\
%  $k_1,k_2,k_3,k_4,k_5$				&1			& Coefficient for discretization of Shield diagram
% 								\\
  $k$			   		&$1$ &  Scheme type, upwind: -1, downwind: 1                        
                                 \\                               
  $m_{cr}$			   &$\text{1}$ & Critical bed slope                         
                                 \\                               
  $N$			    	        &$1$			    &Number of grain-size classes               							\\                              
  $P$					&$1$									&Rouse number
  								\\
  $p$			   &$\text{1}$ & Porosity                         
                                 \\                               
  ${q_x}_{i+1/2,j}$			   &$1$ & Sediment flux of cell surface $(i+1/2,j)$             
                                 \\  
  $q_x$			   &$\text{m}^{3}$ & Sediment flux at x direction                             
                                 \\  
  $q_y$			   &$\text{m}^{3}$ & Sediment flux at y direction                             
                                 \\ 
   $R$			    	        &$1$			    &Reynolds particle number               							\\                                 
    $r_{i+1/2,j}^{\pm}$			   	&$-$ &  Left or right state of Sediment concentration gradient of cell                       
                                 \\
                                 && surface $(i+1/2,j)$
                                 \\                                    
%  $T_{rep}$			    &$s$					&Representative wave period                  					%		\\                               
  $T_s$						&$s$					 & Adaption time
  								\\                                 
  $U$					&$\text{ms}^{-1}$		&Depth-averaged velocity                     							\\
  $U_{rms}$			    &$\text{ms}^{-1}$		&Root mean square velocity                  							\\ 
  $U_{cr}^b$			&$\text{ms}^{-1}$		&Critical velocity for bed load                     				  \\      
  $U_{cr}^s$			&$\text{ms}^{-1}$		&Critical velocity for suspended load                     			  \\
  $U_{i+1/2,j}^{l/r}$			&$\text{ms}^{-1}$ & Left  or right state of depth-averaged velocity of cell surface $(i+1/2,j)$ 
  								\\ 
  $u$ 					&$\text{ms}^{-1}$ 		&x-direction velocity
  								\\ 
%  $u^*_{cr}$			&$\text{ms}^{-1}$		& Critical shear velocity
%  								\\ 
  $u^b_{*,cr}$			&$\text{ms}^{-1}$		& Critical shear velocity for bed load
  								\\ 
  $u^s_{*,cr}$			&$\text{ms}^{-1}$		& Critical shear velocity for suspended load
  								\\                               
  $v$ 					&$\text{ms}^{-1}$		&y-direction velocity  				
  								\\
  $v_{mg}$			        &$\text{ms}^{-1}$		&Velocity magnitude                  							\\                               
  $w_s$ 				&$\text{ms}^{-1}$ 		&Settling velocity of the sediment grain	                       \\
  $w_{s,reduce}$ 		&$\text{ms}^{-1}$ 		&Reduced settling velocity of the sediment grain 
  						         \\ 
  $Z_b$			   &$\text{m}$ & Elevation of sediment surface                             
                                 \\                               
  $z$				    &$m$					&Elevation from sediment bed 
  								\\ 
  $z_0$				    &$m$					&Bottom roughness 
  								\\ 
  $\alpha_{1}$			    	&$1$			    &Settling velocity coefficient                  							\\ 
   $\alpha_{2}$			    	&$1$			    &Settling velocity coefficient                  							\\     
   $\alpha_{i+1/2,j}^{\pm}$	&$1$ & Plus or minus modified parameter of cell surface $(i+1/2,j)$
  								\\
  $\beta_{i+1/2,j}^{l/r}$	&$1$ & Left or right modified parameter of cell surface $(i+1/2,j)$
  								\\
  $\gamma_s$			&$1$									&Ratio between water density and sediment density
								\\
  $\Delta$			    	    &$1$			    &Submerged specific gravity                  							\\  
  $\varepsilon$			   &$1$ &  Order of accuracy                         
                                 \\ 
  $\kappa$				&$1$					 & Von Kaman constant
  								\\                               
  $\nu$					&$\text{m}^{2}{s}^{-1}$					& Kinematic viscosity
  								\\
  $\rho$				&$\text{kg}{m}^{-3}$					& Sea water density
  								\\                         
   $\rho_{water}$				&$\text{kg}{m}^{-3}$ & Sea water without sediment density
  								\\
   $\rho_{s,m}$				&$\text{kg}{m}^{-3}$ &  Sediment density for grain-size class m
  								\\                                                         
  $\Psi_{i-1/2,j}^+$			   	&$-$ &  Right state of flux limiter function of cell surface $(i-1/2,j)$      
                                 \\ 
   $\Psi_{i+1/2,j}^{\pm}$			   	&$-$ &  Left or right state of flux limiter function of cell surface $(i+1/2,j)$
   								\\
   $\Psi_{i+3/2,j}^-$			   	&$-$ &  Left state of flux limiter function of cell surface $(i+3/2,j)$
   								\\                      
\hline
\end{longtable}%}
\end{footnotesize}

\end{appendices}

\newpage
\section*{References}
\bibliographystyle{apa}
\bibliography{model}
\end{document}